\documentclass[aps,preprint,groupeaddress]{revtex4}

\usepackage{amsmath} 
\usepackage{amssymb} 
\usepackage{amsfonts}
\usepackage[dvips]{graphicx} 
\usepackage[]{epsfig} 

\begin{document}

\bibliographystyle{apsrev} 

\title{Salt-specific stability and denaturation of a short salt-bridge forming $\alpha$-helix} 
\author{Joachim Dzubiella} 
\thanks{To whom correspondence should be addressed. E-mail: jdzubiel@ph.tum.de}
\affiliation{Physics Department T37, Technical University  Munich, 85748 Garching, Germany. E-mail: jdzubiel@ph.tum.de }



\begin{abstract}

The structure of a single alanine-based Ace-AEAAAKEAAAKA-Nme peptide in 
explicit aqueous electrolyte solutions (NaCl, KCl, NaI, and KF) at large salt concentrations (3-4 M) is investigated using  $\simeq 1~\mu$s  molecular dynamics (MD) computer simulations. The peptide displays $71 \%$ $\alpha$-helical structure  without salt and destabilizes with the addition of NaCl in agreement with experiments of a somewhat longer version. It is mainly stabilized by direct and indirect (``$i+4$'')EK salt bridges between the Lys and Glu side chains and a concomitant backbone shielding mechanism. NaI is found to be a stronger denaturant than NaCl, while the potassium salts hardly show influence. Investigation of the molecular structures reveals that consistent with recent experiments Na$^{+}$ has a much stronger affinity to side chain carboxylates and backbone carbonyls than K$^{+}$, thereby weakening salt bridges and secondary structure hydrogen bonds.  At the same time the large I$^{-}$ has a considerable affinity to the nonpolar alanine in line with recent observations of a large propensity of I$^{-}$  to adsorb to simple hydrophobes, and thereby 'assists' Na$^{+}$ in its destabilizing action. In the denatured states of the peptide novel long-lived ($10 - 20$~ns) 'loop'-configurations are observed in which single Na$^{+}$-ions and water molecules are hydrogen-bonded to multiple backbone carbonyls. In an attempt to analyze the denaturation behavior within the preferential interaction formalism, we find indeed that for the strongest denaturant, NaI, the protein is least hydrated.  Additionally, a possible indication for protein denaturation might be a preferential solvation of the first  solvation shell of the peptide backbone by the destabilizing cosolute (sodium). The mechanisms found in this work may be of general importance to understand salt effects on protein secondary structure stability and further systematic experiments in this matter are strongly encouraged.

 \end{abstract}


\maketitle 

\section{Introduction}

Identifying and understanding the mechanisms by which salt stabilizes or destabilizes
protein structures is a highly complex issue as, in general, they  result from a subtle 
balance of specific ion-solute and ion-water interactions \cite{baldwin:biophys:96,tobias:science:08}.  
Even the seemingly simpler question how ions interact with planar nonpolar surfaces  and thereby influence their macroscopic interfacial properties is still a matter of ongoing research \cite{jungwirth,kunz,dominik}. It was found for instance that large anions have a surprisingly strong propensity to adsorb to nonpolar surfaces in contrast to the classical continuum picture, where ions are always repelled from low dielectric spaces. 
While for relatively homogeneous solutes (nonpolar atoms, simple colloids, etc.) salt effects can often be at least qualitatively interpreted by rather nonspecific mechanisms such as electrostatic 
screening, changes in solvent surface tension \cite{baldwin:biophys:96}, or dielectric properties~\cite{diel, vila:pnas:00},
the analysis for mixed polar and nonpolar entities on a molecular level, as given by proteins, is far more complicated.  It has been shown recently that even the simple cations sodium and potassium, which are vertical neighbors in the periodic table, exhibit strikingly different behavior in the interaction with protein surfaces, where sodium is strongly favored over potassium \cite{vrbka:pnas:06,uejio:pnas:08}. One  apparent reason is the strong attraction of sodium with protein carboxylates or backbone carbonyls in accord with the 'law of matching water affinities', 
relating ion hydration energies to their respective tendency to form ionic pairs \cite{collins2,collins}.
Similar specifc binding behavior can also be observed for the strong ionic denaturant Guanidinium~\cite{mason}. 
It is, in fact, the distribution of water and ions close to the protein surface, i.e., their  
preferential interaction to (or exclusion from) the latter that intimately relates the change in their chemical potential  and that of the macromolecule.  This useful view has more and more sharpened in the recent years \cite{timasheff:93,parsegian:pnas:00,timasheff:pnas:02,shimizu:jcp:04,ruckenstein,smith} and has been successfully used to analyze salt-specific hydrophobic  association \cite{ghosh:jpcb:05,thomas:jacs:07}.

In particular for large concentrations ($c\gtrsim 1$~M) salt effects are pronounced, 
highly sequence and salt-type specific, and typically lead to changes in protein solubility, stability,
 and/or denaturation that result in the so-called Hofmeister series for the precipitation of proteins~\cite{baldwin:biophys:96}.  Although an order of magnitude higher than at typical physiological conditions ($c \sim 0.1-0.3$~M), 
large salt concentrations play an important biochemical role in the broad field of protein 
crystallization~\cite{dumetz:protein:07}, in food industry as fermentation additives~\cite{dyer}, 
and for the function and stability of biotechnologically interesting halophilic (salt-loving) enzymes~\cite{lanyi}. 
Additionally the  study of protein structures in high salinity solvents is instructive as salt-specific effects are amplified 
and, important from a computational perspective,  can be  sampled more efficiently in molecular dynamics 
(MD) simulations, a typical tool nowadays for the theoretical study of protein folding, 
function, and stability~\cite{andy}.

An ubiquitous and fundamental secondary structure element of proteins is the $\alpha$-helix which is 
stabilized by $(i,i+4)$ backbone hydrogen bonds typically involving four amino acids (aa) per turn. 
The majority of short ($\sim$ 10-20 aa) isolated helices derived from proteins are unstable in solution, unless 
specific side-chain interactions stabilize them. Among these it has been demonstrated that alanine
based peptides have the strongest intrinsic helix propensity~\cite{marqusee:pnas:87,marqusee:pnas:89,spek:jacs:99, chakrabartty:protein:94, scholtz:jacs:91}. Theoretical investigations of the properties
of short helices are vast, either with focus on stability and structure~\cite{sorin:biophys:05,garcia:pnas:02,vila:pnas:00,levy:pnas:01,gnanakaran:proteins:05,fedorov:pccp:07} or folding kinetics \cite{hummer:prl:00, 
monticelli:jpcb:05,wang:jpcb:07}. The action of salt however, has been studied only very 
recently by MD simulations of the trialanine peptides in sodium halide solutions~\cite{fedorov:pccp:07}.
Strongly specific effects on peptide compactness (helicity is ill defined for such a short peptide) and different interface affinities of the ions have been reported in line with the arguments before. 

A very instructive model peptide was experimentally introduced by 
Marqusee and Baldwin \cite{marqusee:pnas:87} who showed with circular dichroism (CD) and other
methods that short (16-17 aa)   alanine-based peptides display high helix propensity when implanted with Glu$^{-}$ and Lys$^{+}$ residues  that possibly form stable salt bridges at pH = 7 (loosely defined, a salt bridge is a weak ionic bond between positively and negatively charged amino acid side chains). A sensitive dependence of helicity on the spacing of the charged groups was detected, i.e., largest helix probability ($\simeq 80\%$ helicity) resulted if glutamic acid (E) and lysine (K) were spaced 4 residues apart as in ``EK($i+4$)'' = Ac-AEAAAKEAAAKEAAAKA-NH$_{2}$, while the reversed order with three spacings, ``KE($i+3$)'' = Ac-AKAAEAKAAEAKAAEA-NH$_{2}$, led to the lowest helicity ($\simeq 20\%$). It was argued that the strong sensitivity may have its origin in the interaction of the charged groups with the electrostatic helix dipole and the larger steric and conformational freedom of the side chains in the ``EK($i+4$)''  case allowing for a higher salt bridge forming potential. A quantification and  molecular details of these hypotheses however, remained elusive. Importantly, it was shown that  addition of a simple monovalent salt like NaCl can monotonically destabilize the  ``EK($i+4$)'' helix with increasing  concentration ($\simeq 57 \%$ at 3 M) at pH = 7, which could not be interpreted by screening only \cite{marqusee:pnas:87}. 

Further work on other peptides revealed that the stabilizing potential of salt bridges strongly depends on sequence context and suggested that they can contribute significantly to thermal stability, obviously an appealing issue for protein engineers~\cite{mak,olson:proteins:01}. In particular, in a recent MD study of the thermal stability of a 20 aa long ``EK($i+3$)'' peptide it was indeed found that salt bridges  contribute to thermal stability but not significantly~\cite{ghosh:biophys:03}.  Another, possibly more important  stabilizing mechanism was identified, where the relatively long Lys side chain shielded  specific backbone hydrogen  bonds  from the direct interaction with water. This screening mechanism by long side chains has been observed also for other peptides to act stabilizing on the helical structure \cite{vila:pnas:00, garcia:pnas:02}. 

In this work we employ standard all-atom  MD simulations to investigate the structure and (helix) stability of a shorter version of the salt-bridge forming ``EK($i+4$)'' peptide with 12 aa and sequence Ace - AEAAAKEAAAKA-Nme.  We  systematically explore the (de)stabilizing action of four different salts (NaCl, NaI, KCl, KF) at large concentrations ($c \gtrsim 3$ M) and analyze the molecular and structural details in relatively long $1~\mu$s runs. Particular focus is given to peptide  structure, role of salt bridges and backbone shielding, and the specific influence of solvent and ions thereon.  An attempt is made to interpret peptide (de)stabilization by a preferential hydration/exclusion mechanism of solvent and ions.

\section{Methods and systems}

MD simulations are performed using the parallel module {\it sander.MPI} in the simulation package Amber9.0 with the ff03 force-field for the peptides and the rigid and nonpolarizable TIP3P  water for the 
solvent~\cite{amber}. Parallel computing on 8-16 processors is executed on the supercomputer HLRBII of the Leibniz-Rechenzentrum M\"unchen (LRZ). All simulated systems are maintained at a fixed pressure of $p=1$ bar and a temperature $T$ by coupling to a Berendsen barostat and Langevin thermostat~\cite{amber}, respectively. The cubic and periodically repeated simulation box has edge lengths $L~\simeq 36$~\AA~ including 
approximately 1500 water molecules. Electrostatic interactions are  calculated by particle mesh Ewald summation and all real-space interactions (electrostatic and  van der Waals) have a cut-off of 9~\AA. The peptide is generated using the {\it tleap} tool in the Amber package~\cite{amber}. 

Cations and anions in our study are modeled as nonpolarizable  
Lennard-Jones spheres with charge and interaction parameters as supplied by Dang~\cite{dang} and 
summarized in Tab.~I. The considered concentrations result from $90$ ion pairs in the simulation box.  
We investigate the structure and helix stability of the 12 aa long peptide with the acetyl (Ace) and amine (Nme) capped sequence Ace-AEAAAKEAAAKA-Nme  forming three $\alpha$-helical turns in the fully folded state,  where Glu2 and Lys6, and Glu7 and Lys11, are potentially able to form a salt bridge, respectively.  We simulate at a temperature $T=274$~K, relevant for comparison to experimental  trends for the somewhat longer (17 aa) analog in~\cite{marqusee:pnas:87} in NaCl. The influence on helix stability of large concentrations ($ \gtrsim 3.0$ M) of the simple  monovalent salts NaCl, KCl, NaI, and KF is investigated. We simulate each system for $1~\mu$s requiring  a total simulation  time of $5~\mu$s.  All simulation snapshots are visualized using VMD \cite{vmd}.  Trajectory analysis is done using the ptraj  tool  in the Amber package \cite{amber}, where, in particular, secondary structure elements such as the helicity (i.e., helix percentage per peptide configuration) are identified using the DSSP method by Kabsch and Sander \cite{dssp}. In our MD we start with initially unfolded peptide configurations. Even the most  stable helices feature typical folding times of at least  25-50 ns before the first folding event, an order of magnitude slower than in implicit solvent calculations of the 17 aa  long ``EK($i+4$)''peptide \cite{wang:jpcb:07}. We simulate  further $\simeq 50$ ns before gathering statistics.

We attempt to examine and interpret some of our data in the framework of preferential interactions 
\cite{timasheff:93}. Briefly, the excess chemical potentials of a ternary mixture of solute (macromolecule)  $\mu_{M}$, cosolute (salt) $\mu_{s}$, and water $\mu_{w}$ are intimately connected 
by Gibbs-Duhem-like relationships:  ${\rm d}\mu_{M}^{ex} = -\Gamma_{Ms} {\rm d}\mu_{s}$ and ${\rm d}\mu_{M}^{ex} = -\Gamma_{Mw} {\rm d}\mu_{w}$.  The experimentally accessible  preferential  interaction parameters are defined by 
\begin{equation}
\Gamma := \Gamma_{Ms} = N_{ s}\left( 1-\frac{N_{ w}/N_{ s}}{n_{ w}/n_{ s}} \right),\;\;\;\;\;\;
\Gamma_{Mw} = N_{ w}\left( 1-\frac{N_{ s}/N_{ w}}{n_{ s}/n_{ w}} \right), 
\end{equation}
where $n_{ s}/n_{ w}$ is the ratio of the number of cosolutes and water molecules
in the bulk, and $N_{ s}/N_{ w}$ is the ratio of the number of cosolutes and water molecules 
in a local domain around the macromolecule.  Thus, the excess chemical potential of the macromolecule's
configuration is directly related to the distribution of cosolutes  in the solute's vicinity. If the cosolutes are excluded (preferential hydration) from the macromolecule's local domain,  then $\Gamma < 0$.  If cosolutes bind to the macromolecule, then $\Gamma > 0$ (preferential dehydration). It is observed empirically that a strong preferential hydration (cosolute exclusion) typically stabilizes protein structures although it can not be used as a stability criterion, i.e., $\Gamma < 0$ does not necessarily imply the stabilization of the native structure~\cite{arakawa}. Whether the preferential interaction framework can be used as a criterion for protein stability is therefore questionable. 
The microscopic reason is fine balance between exclusion of the cosolute from and its specific binding to the protein surface in dependence of the chemical nature of the latter~\cite{arakawa,schellman}.  These favorable binding events are typically considered responsible for protein denaturation but it is in general unclear where the particular binding sites are or who the binding partner is. The detailed molecular origins that define $\Gamma$ are thus not 
well understood. 

\section{Results and discussion}

First of all, a comment should be made on the reliability of classical molecular force-fields used 
in MD simulations. We employ  ion parameters that were introduced by Dang \cite{dang} as the 
default Amber parameters, e.g., Na$^{+}$ and K$^{+}$ from \AA qvist~\cite{aqvist} mixed with Cl$^-$ from Dang,
are known to be faulty at larger salt concentrations. They show a too strong 
ion pairing propensity and premature freezing at concentrations considerably lower than saturation~\cite{auffinger:jctc:07,suk}. We have tested the Dang parameters for a wide range of densities in the homogeneous case and always found a reasonable fluid structure as can be judged by looking at the radial distribution 
functions~\cite{auffinger:jctc:07}. Their dependence on the particular water model has been found to be 
weak~\cite{auffinger:jctc:07, patra:jcc:04}. However, macroscopic bulk properties such 
as the osmotic pressure for these force fields at large concentrations however, have not been 
systematically investigated to the best of our knowledge.  The necessity of the explicit consideration of the 
polarizability of ions is still a matter of debate~\cite{jungwirth,kunz}.   
Regarding the water model we use TIP3P parameters as it is the default water model in Amber and 
hopefully reasonably balanced with the solute force-fields, although it is known that other models 
(such as SPC/E) seem to perform better in describing water bulk properties \cite{water}.
Also intrapeptide potentials are typically not free from error when benchmarked to experiments or other more reliable calculations, see, e.g. references~\cite{sorin:biophys:05, hornak:proteins:06, gnanakaran:proteins:05}. Therefore, we do not  claim to be quantitative in our work but focus on the discussion of effects and qualitative trends. 

\subsection{Helix stability}

Examples for peptide helicity vs. simulation time $t$ are plotted  in Fig.~1 for a) the salt-free case, b) 3.7 M  NaCl, and c) 3.5 M KCl. Without salt the helicity can be considerable ($\sim 80\%-90\%$) but two large-scale unfolding transitions are observed at $ t \simeq 130$ and 900~ns, where the peptide is basically unfolded (helicity $<$ 50\%) for $\sim$ 50 ns. The total helicity averaged over the whole trajectory is found to be $71 \%$, not far from the experimental value of about $80\%$ of  the 17 aa long analog~\cite{marqusee:pnas:87}.   Adding NaCl at a large concentration clearly leads to destabilization of the helical structure as shown in Fig.~1 b). States with more than $50\%$ helicity are rarely sampled and stable only on $\sim$ 50-200 ns time periods. The average helicity goes down to 37\% for $c =$ 3.7~M.  The destabilizing trend is in agreement with experimental  measurements of the longer analog (17 aa), where a continuously decreasing helicity with increasing NaCl concentrations was found with a value of 57\% at $c \simeq 3$ M \cite{marqusee:pnas:87}.  In contrast to NaCl, we find only a slight destabilizing action when replacing sodium by the 30\% bigger potassium ion (cf. Tab.~I) as shown in Fig.~1 c). The average helicity of the system of 63\% is close to the salt-free case even at a large concentration of $c = 3.5$ M KCl. The average helicity of all investigated systems is summarized in Tab. II. 

The helicity resolved by individual peptide residues for the same systems is plotted in Fig.~2. The helicity distribution for the salt-free case and 3.7~M KCl is highest in the center of  the peptide and diminishes to the terminal ends as found before in studies of a similar peptide involving salt bridges~\cite{ghosh:biophys:03}.  Note that these 'finite-size' or end-effects lead to a somewhat smaller total helicity in our MD study when compared to the experimental, longer analog with 17 aa. Further inspections of the data reveals that the helicity along the peptide is asymmetric with respect to the terminal ends, i.e., there is a distinct jump in helicity between the Glu2 and Ala3 residues.  The reason is the  preferable interaction of the carboxylate group of  Glu2 with the  N-terminus  which was also observed in  an implicit solvent study of the same sequence pattern \cite{wang:jpcb:07}.  It is known experimentally that  'N-capping' by specific side chains strongly influences the helix distribution along a given peptide~\cite{vasquez:biopolymers:87,serrano:nature:89,chakrabartty:pnas:93}. Although negatively charged residues close to the N-terminus have been found to stabilize the helix locally by a preferential charge - helix dipole interaction we observe  that Glu2 can be dragged to the N-terminus enabling Lys6 to interact with the Ala1 or Glu2 carbonyls, thereby destabilizing the Glu2-Lys6 salt-bridge and their backbone hydrogen bonds (leading to a local destabilization of the helix). We will return to this issue in the next section where the role of salt bridge formation is discussed in more detail.  With 3.7 M NaCl 
the helicity per residue clearly diminishes for each residue as is also illustrated in Fig.~2. 

For a better understanding of the equilibrium structure distribution of the investigated peptide it is instructive to calculate the probability $P({\rm rmsd})$ by which a configuration root mean square deviates (rmsd) from the fully helical reference structure.  The latter was chosen randomly from configurations with  100\% helicity. We found that the following results did only slightly ($\lesssim 5\%$)  depend on its particular choice. The according free energy  $\Delta G({\rm rmsd}) = -k_{B}T \ln(P)$ vs. the rmsd reaction coordinate without salt,  NaCl, and KCl is shown in Fig.~3. At zero salt concentration and with 3.7 M KCl  there are two distinct minima at which the peptide is in the fully helical state with three helical turns [rmsd $\simeq 1.6$~\AA,  cf.~Fig.~3~a)],  or features two neighboring helical turns [rmsd $\simeq 3.1$~\AA, cf.~Fig.~3~b)]. The free energy barrier along our chosen reaction coordinate between these two states is about $1.5 k_{B}T$. In the salt-free case larger rmsd appear only with a much higher energy penalty, while for KCl also two small local minima are present at $5.25$~\AA~ and  6.5~\AA. The situation is qualitatively different with 3.7 M NaCl. The minimum corresponding to the full helical state disappears and a  broad minimum occurs at larger rmsd values $\sim 5-7$~\AA.  Detailed inspection of our MD data reveals  that this broad minimum is occupied by mainly two distinct configurations, which can obviously not be resolved with our simple rmsd reaction  coordinate. At a rmsd $\simeq 5.5$~\AA~ we observe long-lived configurations that feature two helical turns at the both ends of the peptide with an intermediate loop, see the snapshot in Fig.~3~c), while the rmsd $6.5-7$~\AA~ mainly represents the coil state where helicity completely vanishes, see Fig.~3~d). The ``loop''-states [Fig.~3~c)] in NaCl are long-lived and stabilized by {\it specific} sodium and water binding to the peptide backbone, an intriguing observation we will discuss later in more detail.

In order to test the influence of changing the anion we replaced NaCl by NaI and KCl by KF. We observed 
only little change in the results above for the potassium salts, but a noticeably larger destabilizing action for 
the iodide salt.  In fact, NaI is known to be a strong denaturant \cite{baldwin:biophys:96}.
It seems however,  that for the particular peptide and ions studied the difference in the cation charge density is the crucial parameter which determines the denaturation capability of the salt. Although comparable in size and often observed to be similar in their salting-in and salting-out actions on proteins \cite{baldwin:biophys:96,nandi:jacs:72a,nandi:jacs:72b}, qualitatively different influence on the solubility of amino acids \cite{khoshkbarchi:ind:97}, on the activity  and stability of halophilic proteins \cite{madern:extrem:00}, and a striking difference in binding to  protein carboxylates and carbonyls \cite{vrbka:pnas:06,uejio:pnas:08} have been observed experimentally. Unfortunately, no experimental data on KCl, KF,  or NaI action on the  ``EK($i+4$)'' peptide for a more meaningful and verifying/falsifying comparison is available yet. 

\subsection{Salt bridge formation and backbone shielding}

The charged residues Glu2$^{-}$ and Lys6$^{+}$, and residues Glu7$^{-}$ and Lys11$^{+}$ are potentially able to form EK-type salt bridges, which may directly stabilize the helix by shifting  the equilibrium to the helical, more compact state \cite{marqusee:pnas:87} and, additionally,  may contribute to helix stability through backbone desolvation and shielding from water  \cite{vila:pnas:00, garcia:pnas:02,ghosh:biophys:03}. To elucidate these mechanisms in more detail we calculate the (normalized) probability distribution $P(r)$, where $r$ is the distance  between the carboxylate carbon atoms on Glu and the amine nitrogen on Lys for the first (Glu2-Lys6) and second (Glu7-Lys11) salt bridge, averaged over the whole MD trajectory.  Examples for  the salt-free case and large NaCl and KCl concentrations are shown in Fig.~4 for the  Glu7-Lys11 pair. The first (contact) peak  for distances $r_{d}\lesssim 4.3$~\AA~ indicates a {\it direct} salt-bridge while the  second peak $r_{d}\lesssim r \lesssim r_{id}\simeq 7.0$~\AA~corresponds to  an {\it indirect} -- by one water molecule separated -- salt bridge. Illustrating MD snapshots are shown at the top of Fig.~4. With those distribution at hand we  estimate the probability $p(r_{i}) = \int_{0}^{r_{i}} {\rm d}r P(r)$, $i=d,id$,  of finding a direct or both, direct {\it and} indirect salt bridges in the course of our simulation by integrating up to $r_{d}$ or $r_{id}$, respectively. In the salt-free case only by $p(r_{d})=11\%$ of the time a direct salt bridge is formed by one of the Glu-Lys pairs. This value  is in accord with the results of Ghosh {\it et al.}~\cite{ghosh:biophys:03} on the similar ``EK($i+3$)'' peptide who found that possible EK salt bridges spend only $10-20\%$ in contact configuration and thus probably do not play an immediate role in helix stabilization.  Both, direct {\it and} indirect salt bridges however, have a higher probability  $p(r_{id})=$ 27\% for the Glu2-Lys6 pair and in $P(r_{id})= 43\%$ for the Glu7-Lys11 pair. As discussed previously the asymmetric stability behavior between the two salt bridges is the consequence of the interaction of Glu2 with the N-terminal. The probability values for forming direct and both, direct and indirect salt bridges for all investigated  systems are  summarized in Tab.~II. 

The distribution function depends on salt type and concentration as is also illustrated in Fig. 4. For NaCl  
the salt bridge formation probability is decreased by more than 50\%. Obviously this strongly 
correlates with the average helicity in the system as in a non-helical state the charged side chains are too 
far apart to come into a contact state. For the large KCl concentration the distribution changes
much less when compared to NaCl. For further analysis, in Fig. 5 
we plot the radial distribution function (rdf) of water and cations (anions) around the Glu (Lys) headgroup, i.e., 
the Glu carboxylate carbon (Lys side chain nitrogen). Strikingly, and observable from the contact peak, there is strong affinity of sodium to the carboxylates over potassium as has been found in previous experimental studies and
quantum mechanical calculations \cite{vrbka:pnas:06,uejio:pnas:08}. The rdf of potassium on the other hand
is comparable in magnitude to that of the water oxygen;  to quantify, we estimate the coordination 
number of species $i$ in the first carboxylate solvation shell with
\begin{equation}
N_{i}(r_{c}) = c_{i}\int_{0}^{r_{c}}{\rm d}^{3}r g(r),
\end{equation} 
where $r_{c}=4.3$\AA~ is chosen to be the extension of the first solvation shell and $c_{i}$ is the 
number density of species $i$. While in the salt-free case 8-9 water molecules directly solvate the carboxylate, $\sim 2$ of them are replaced on average by sodium ions in the NaCl solution, indicating a strong specific binding of sodium. Note also the decreased contact value of the water-carboxylate rdf in the NaCl solution in Fig.~5. In KCl we find on average only $\sim 0.5$ ions which replace water molecules, suggesting that potassium is  much weaker in breaking direct or indirect salt bridges. In contrast, sodium directly competes with water and Lys-nitrogen atoms in binding to the carboxylates and has a strong impact on salt bridge formation. The distributions of water and anions around the Lys nitrogen is plotted in the inset to Fig. 5:  F$^{-}>$ Cl$^{-}>$ I$^{-}$ is the preferred sequence of interaction as could have been expected from electrostatic considerations, i.e., there is a stronger attraction with larger ion charge density.  Calculating coordinations numbers as above however, reveal that also  only on average $\sim 0.5$ fluoride ions are able to replace a water molecule in the water solvation shell, still much less effective in replacing water than sodium around the carboxylates.

Another secondary-structure stabilizing effect has been proposed to be the local or nonlocal shielding of backbone hydrogen bonds from water by large side chains~\cite{vila:pnas:00, garcia:pnas:02,ghosh:biophys:03}. For an analysis we follow Ghosh {\it et al.}~\cite{ghosh:biophys:03} and plot in the inset to Fig.~4 the average number of water molecules in the first solvation shell of the  backbone carbonyls resolved by amino acid residue. In the systems with large average helicity (salt-free and KCl) the backbone carbonyls 2 or 3, 6,  7 are distinctively less solvated (1.1-1.3 water molecules) than the others. These carbonyls indeed belong to or are in immediate vicinity of the two Glu and the central Lys side chains. By salt-bridging along the folded peptide, exactly their carbonyls are screened from the surrounding water and the hydrogen bond stability of the latter is very likely to be increased.  Snapshots are shown at the top of Fig.~4, illustrating the screening effect of the Glu7-Lys11 direct and indirect salt bridges.   From the inset to Fig.~4, we clearly see that adding salt leads  to a larger hydration of the backbone, probably due to the induced unfolding and its exposure to the solvent. Interestingly, even for the large NaCl concentration salt bridge formation is not entirely suppressed, see Tab. II, and backbone carbonyl 7 remains least solvated. 

An interesting conclusion from this analysis is that direct {\it and} indirect salt bridges  -- while they may not play a dominant {\it immediate} role -- stabilize the helix by efficiently screening some of their backbone hydrogen bonds. The strong affinity of sodium over potassium to the carboxylates makes the  former
a powerful competitor to water in the binding to the Glu side chains, thereby decreasing the probability
of forming direct or indirect salt bridges. 

\subsection{Preferential hydration and ion binding}

In order to check whether similar specific binding events as found in III.B. can also be detected 
for other parts of the peptide, we have calculated the rdfs  between ions and 
the backbone amide oxygens and nitrogen atoms, averaged over the whole trajectories. Examples for the 
cation and anions  are plotted in Fig. 6 a) and b), respectively.  Analogous to the results in III.B.  a strong 
attraction of sodium to the carbonyl oxygens is observed in striking contrast to a weak affinity of potassium. The anions however,  interact with the backbone even weaker, cf. Fig. 6 b).  Interestingly, iodide shows here the highest first peak  compared to the other anions (I$^{-}> $ Cl$^{-}>$ F$^{-}$) conversely to its interaction with the Lys headgroup, see the inset of Fig.~5, where (F$^{-}>  $ Cl$^{-}>$ I$^{-}$).  Related to this, the strong affinity of sodium to the backbone is enhanced in the presence of iodide. We argue that these effects may have two  possible origins: firstly, the bulk activity of NaI is larger than that of NaCl at  the same concentration \cite{activity},  in other words, transferring a sodium-iodide pair from a NaI solution to a reference solution costs less energy than from NaCl.  Secondly, it was shown that  the relatively large anions have an affinity to nonpolar surfaces \cite{jungwirth,kunz,dominik} or hydrophobic solutes~\cite{kalra}.  To inspect this hypothesis for our (partly nonpolar) peptide, we plot the rdf between the carbon atom in the alanine side chain and  cations or anions in the insets to Fig. 6 a) and b), respectively. Indeed we  find that Iodide has the strongest affinity to the nonpolar side chains from all considered ions, giving rise to a relatively high affinity to the protein surface.  These findings agree with MD of a toy model of a heterogeneous biomolecule, 
where it has been  argued and demonstrated that fluoride and iodide binding strongly depends on the local surface characteristics~\cite{lund:langmuir:08}. They also support the perspective that the iodide propensity  to nonpolar surfaces may indeed impact protein stability \cite{jungwirth,kunz}, a mechanism also proposed for other large ions such 
as Guanidium~\cite{mason}.

In the following we attempt to examine and interpret this data in the framework of preferential interactions
(see Methods).  To look at the ionic distribution in detail we separately calculate the preferential interaction parameter $\Gamma_{i}$ for every ionic species $i=\pm$, so that $\Gamma =\sum_{i=\pm} \Gamma_{i}$.
We define the local peptide domain using an appropriate cut-off $r_{c}$
 around every atom for counting water and salt particles. We find that our results do qualitatively not depend on  the choice of $r_{c}$, for which we tested values between 4 and 8~\AA. In Fig. 7 we plot the results 
for $r_{c}=4.3$~\AA~ -- which roughly corresponds to the water first solvation shell (Figs. 5 and 6)  
-- versus peptide helicity. We make following observations: firstly, $\Gamma_{i} < 0$ for all ions, showing preferential hydration of the peptide in all configurations and for all salts. Secondly, in agreement with the preferential interaction picture the destabilizing salts show the largest affinity to  the peptide, indeed in the order of their destabilizing action. Thirdly, the dependence of $\Gamma$ on the helicity is surprisingly weak given  the broad conformational changes of the peptide and shows hardly structure. For NaI and NaCl,  $\Gamma$ increases by $\sim$ 30\% and 6\%, respectively,  when going from the  helical to the coil state,  while for the other salts the change is less than 5\%. Our data gives thereby microscopic support of the  statement that preferential hydration  can not be used as a criterion for protein stability~\cite{arakawa}.   Overall, the  large iodide has a  surprisingly strong affinity to the protein, obviously due to the more attractive interaction with the nonpolar Ala side chains and backbone nitrogen atoms when compared to the other anions  or potassium, see the discussion of Fig. 6 above. 

It seems that the denaturation of the helical structure by NaCl and NaI  can be attributed to 
the favorable interaction of sodium with the backbone carbonyls, amplified by weak anions
such as iodide.  Therefore it is instructive to calculate the number of ions and the 
preferential interaction parameter just for the  first water solvation shell of the {\it protein 
backbone} which we define as $\Gamma^{\rm bb}$, evaluated for every ionic species.
From the the first minimum of the backbone-water rdfs in Fig.~6 we estimate the
extension of the first solvation shell to be $r_{c} = 3.5$~\AA, which we now take to 
define the local domain around the backbone. Results for the number of ions in the first backbone solvation shell are plotted in Fig. 8:  while  less than one potassium or anionic particle can be found on average close to the backbone, typically  more than one sodium interacts with it, strongly depending on helicity. In the nonhelical state approximately 2 and 2.5 sodium ions are on average bound to the backbone for NaCl and NaI, respectively. If these numbers are translated into $\Gamma^{\rm bb}$, see the inset of Fig.~8, we observe that  $\Gamma^{\rm bb}$ changes sign for the nonhelical states for a helicity $< 55 \%$ and
$<75 \%$ for NaCl and NaI, respectively. Potassium (shown) and anions (not shown) do not 
exhibit this behavior. If $\Gamma^{bb}$ is calculated including the second solvation shell
($r_{c} \simeq 6.5$\AA) a change of sign for sodium is not observed either. On the basis of
this data one could speculate that preferential solvation only of the first backbone solvation 
shell , $\Gamma^{\rm bb}$(1st solv. shell)$ > 0$, by the destabilizing species (here sodium) 
may be a criterion for protein denaturation. 

Finally, inspection of MD trajectories for NaCl and NaI reveal that the strong interaction of 
sodium with the backbone carbonyls can result in intriguingly long-lived protein configurations where 
sodium is bound and actively  involved in the protein structure, as shown in Fig. 9. 
The central part of the peptide loops around a single sodium ion, thereby binding it with 
3-4 backbone oxygens, while still a partly helical structure  can be maintained, cf. Fig. 9 a). 
Sometimes an additional  water molecule is captured by the backbone-ion complex and 
binds, as illustrated in Fig. 9 b) and c). These states are surprisingly stable on a long 
10-20~ns  time scale. We have not observed such long-lived states involving 
potassium or anions. 

\section{Concluding remarks}

In summary, by using all-atom MD simulations we have provided molecular insights into 
the structural stability of a short salt-bridge forming peptide. We find that 
specific structural mechanisms such as salt bridges and side chain shielding can 
stabilize a helical structure in accord with experiments. Although direct salt bridges 
are  found only rarely, the action of indirect (by one water molecule separated)
salt bridges must not be overlooked as they may contribute significantly to the 
shielding of local backbone hydrogen bonds. These stabilizing mechanisms are observed to be 
suppressed by the  specific binding  of sodium to  carboxylates and backbone carbonyls, thereby 
shifting the equilibrium from  helical to coil states. This specific binding was not observed for the  
$\simeq 30 \%$ larger potassium ion.  

A surprisingly large affinity to the peptide for iodide is observed however, in particular
to the nonpolar, hydrophobic parts of the protein, and as previously observed in studies 
and experiments of large anions at  planar hydrophobic interfaces~\cite{jungwirth,kunz,dominik}, 
toy model proteins~\cite{lund:langmuir:08}, and simple hydrophobes~\cite{kalra}. The reasons are probably a mix of a larger bulk activity for halide salts with larger anions \cite{activity} and a specific water-assisted
affinity to nonpolar residues what renders iodide itself a rather hydrophobic entity. 
Bulk effects and the high peptide affinity of iodide leads to an increased number
of sodium ions close to the peptide, implying that iodide directly and indirectly increases 
peptide dehydration and thus promotes helix destabilization. These findings support the 
perspective that the iodide propensity to nonpolar surfaces may impact protein stability \cite{jungwirth,kunz}.
They also highlight the highly synergetic action of the strongly electrostatically coupled 
co- and counterions, which can be attracted to the protein surface for different reasons.

In an attempt to interpret our findings in the preferential interaction framework 
we find indeed that for the strongest denaturant, NaI, the peptide is least 
hydrated and $\Gamma$ is maximal. It remains negative however, showing a preferential 
hydration. Resolving $\Gamma$ vs. helicity shows weak structure, possibly due to canceling contributions from 
polar and nonpolar parts of the proteins and has to be more thoroughly investigated.
However, if $\Gamma$ is calculated for the destabilizing species (sodium) in the 
first backbone solvation shell only,  a strong slope and a change of sign is 
observed at intermediate helicities. This preferential solvation of the backbone 
[$\Gamma_{\rm Na^{+}}^{\rm bb}$(1st solv. shell)$>0$] may  serve as an criterion 
for protein denaturation.  An experimental verification of  this hypotheses is probably hardly 
feasible but it may provide food for thought and inspire  further (computational) investigation 
on this matter as the microscopic origins of  $\Gamma$ are subtle. 

The mechanism found may be of general importance to understand cosolute effects on protein 
secondary structure stability and further experiments probing systematically the salt-specific 
action on the $\alpha$-helical stability of this  or similar short peptide are strongly encouraged. 
As we have shown in this work, molecular insights from MD simulations can provide valuable information 
to understand the intricate mechanisms in solvent-protein interactions and thereby protein stability and folding.  For this however, accurate MD force-fields, in particular for ions at moderate to large 
concentrations, are an essential prerequisite to avoid possible artifacts~\cite{auffinger:jctc:07, suk}, 
and need to be benchmarked to experiments. Finally, we note that the novel long-lived 
'loop'-configurations in the denatured/unfolded states in which sodium and water is bound and immobilized by the peptide  backbone may be experimentally accessible by nuclear magnetic relaxation dispersion 
methods (NMRD)~\cite{denisov:nature} or time-resolved FRET measurement~\cite{kiefhaber}. 

\section{Acknowledgements}

  J. D. is grateful to D. Horinek and I. Kalcher for useful discussions, the Deutsche 
  Forschungsgemeinschaft (DFG) for support within the Emmy-Noether-Program, 
  and the LRZ M\"unchen for computing time on HLRBII.

  \clearpage

\begin{table}[h]
  \caption{Atom-atom Lennard-Jones parameters $\sigma$ and $\epsilon$ and charge $q$ used in our study for the ion ~\cite{dang} and TIP3P water oxygen (O) and hydrogen (H) interactions.}
\begin{center}
\begin{tabular}{| c | c | c | c | }
\hline
  atom &  $\sigma/$\AA & $\epsilon/$(kJ/mol)  &  charge $q/e$  \\  
    \hline
  Na$^{+}$  & 2.584 & 0.418 & +1 \\
  K$^{+}$  & 3.332 & 0.418 & +1 \\
  F$^{-}$  & 3.120 & 0.753 & -1 \\
  Cl$^{-}$  & 4.401 & 0.418 & -1 \\
  I$^{-}$  & 5.171 & 0.418 & -1 \\
  \hline
  O  & 3.151 & 0.636 & -0.834 \\
  H  & 0 & 0 & +0.417 \\ 
  \hline
\end{tabular}
\label{tab1}
\end{center}
\end{table}

\begin{table}[h]
  \caption{Helicity and salt bridge probabilities of the peptide for different types of salt at
  concentration $c$. $p_{\rm sb1}(r)$ and $p_{\rm sb2}(r)$ correspond
  to the probability of finding the Glu2-Lys6 and Glu7-Lys11 head groups within a distance $r$, 
   where we define $r=4.3$~\AA~  and $r=7.0$~\AA~ to be direct and indirect salt bridges, respectively.}  
\begin{center}
\begin{tabular}{| c | c | c ||| c | c || c| c|}
\hline
  salt &  $c$/M & helicity &  $p_{\rm sb1}(4.3{\rm \AA})$ & $p_{\rm sb1}(7.0{\rm \AA})$  & $P_{\rm sb2}(4.3{\rm \AA})$ & $P_{\rm sb2}(7.0{\rm \AA})$   \\  
    \hline
  --      & 0.0 & 0.71  & 0.10 & 0.27 & 0.11 & 0.43 \\
  NaCl & 3.7 & 0.30  & 0.03  & 0.07 & 0.06 & 0.18 \\
  KCl & 3.5 & 0.63   &  0.02  & 0.11& 0.10 & 0.33 \\
  NaI   & 3.3 & 0.23 & 0.04  & 0.10& 0.09 & 0.24  \\
  KF    & 4.0 & 0.58 & 0.03  & 0.16& 0.09 & 0.31 \\
\hline 
\end{tabular}
\label{tab1}
\end{center}
\end{table}

\clearpage 

\begin{figure}[htp]
\includegraphics[width=14cm,angle=0]{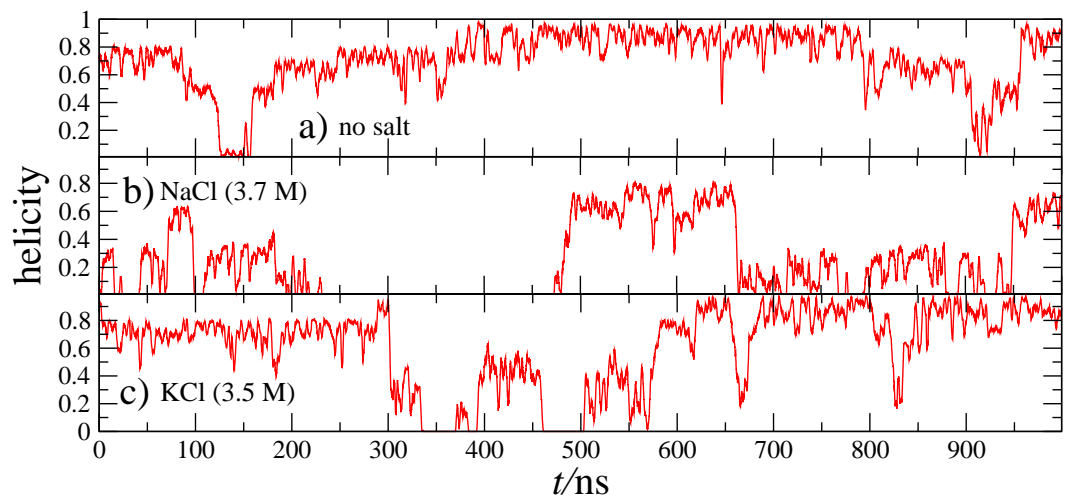}
\caption{Peptide helicity vs. time $t$ for  a) no salt, b) $c=$ 3.7 M NaCl, and 
c) $c=$ 3.5 KCl. Local time averaging over a 2 ns window of the raw MD data has been performed.}
\label{fig:1}
\end{figure}

\begin{figure}[htp]
\includegraphics[width=14cm,angle=0]{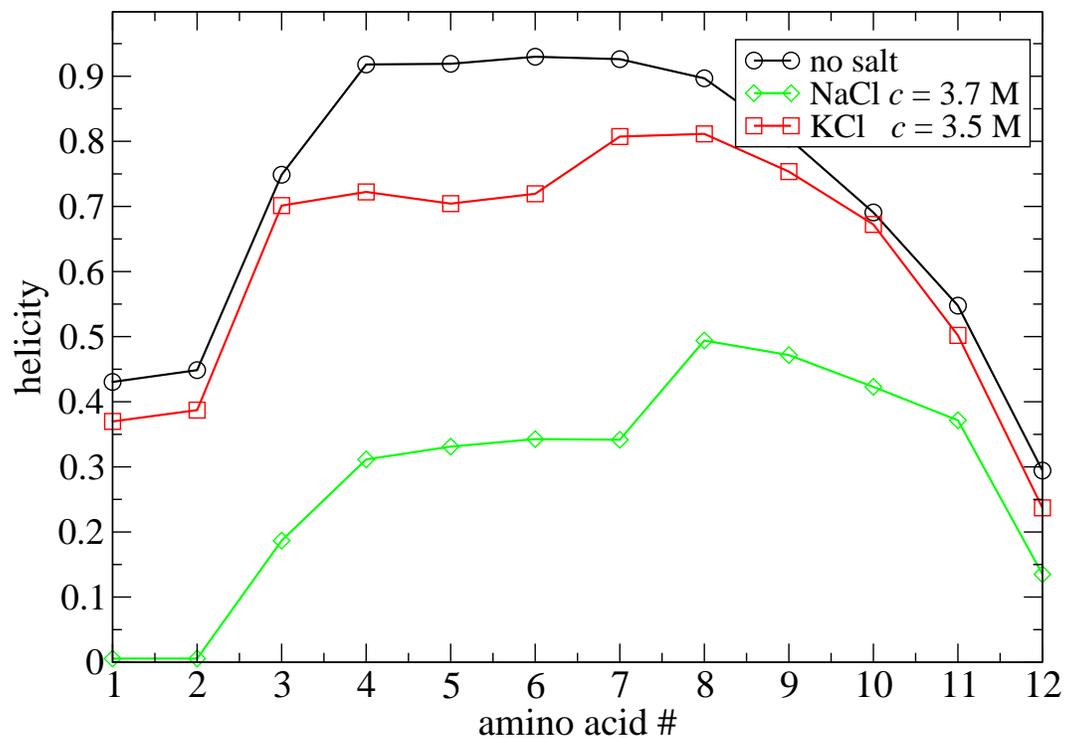}
\caption{Average helix probability (helicity) of each residue in the peptide for no salt and large 
 NaCl and KCl concentrations. Lines are guide to the eye.}
\label{fig:1}
\end{figure}

\begin{figure}[htp]
\includegraphics[width=14cm,angle=0]{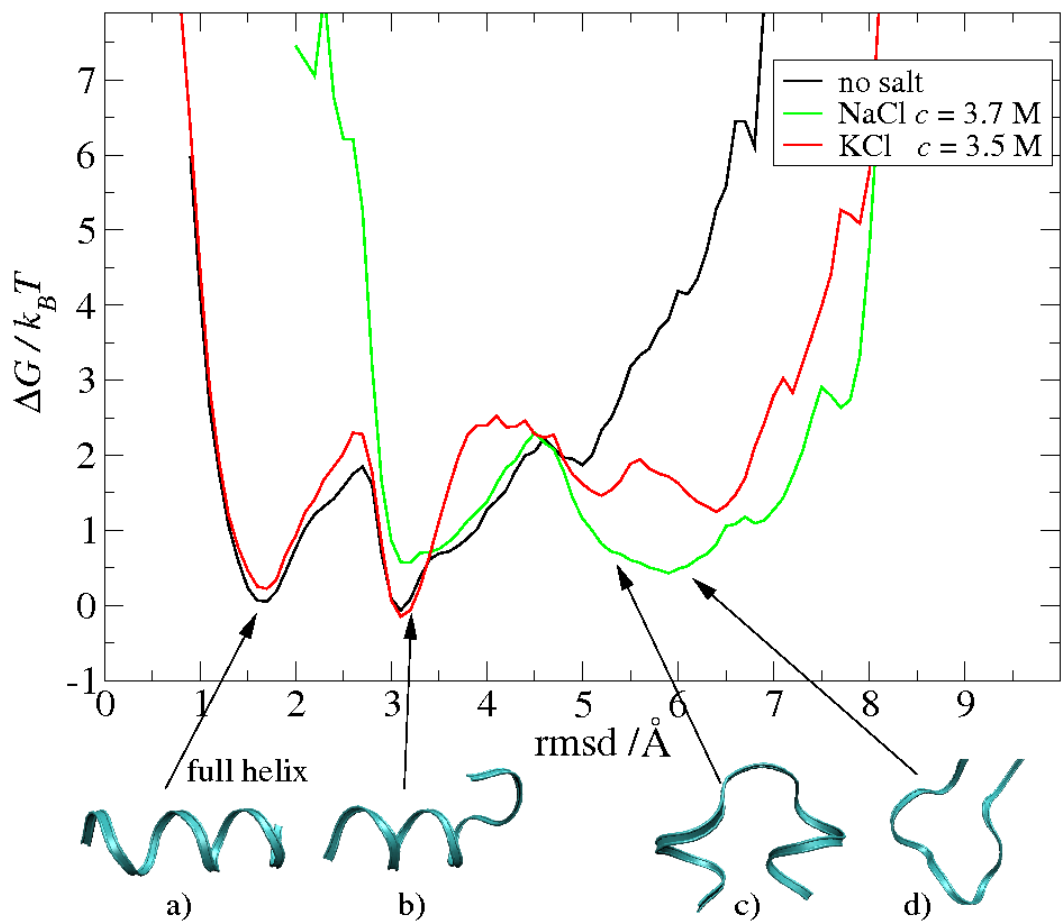}
\caption{Free energy along the rmsd (root mean square deviation from a 'perfect' helix formed by this
peptide) 
reaction coordinate. The MD snapshots of the peptide backbone at the bottom exemplify configurations with high probability (low free energy) with an rmsd location indicated by the arrows. }
\label{fig:1}
\end{figure}

\begin{figure}[htp]
\includegraphics[width=14cm,angle=0]{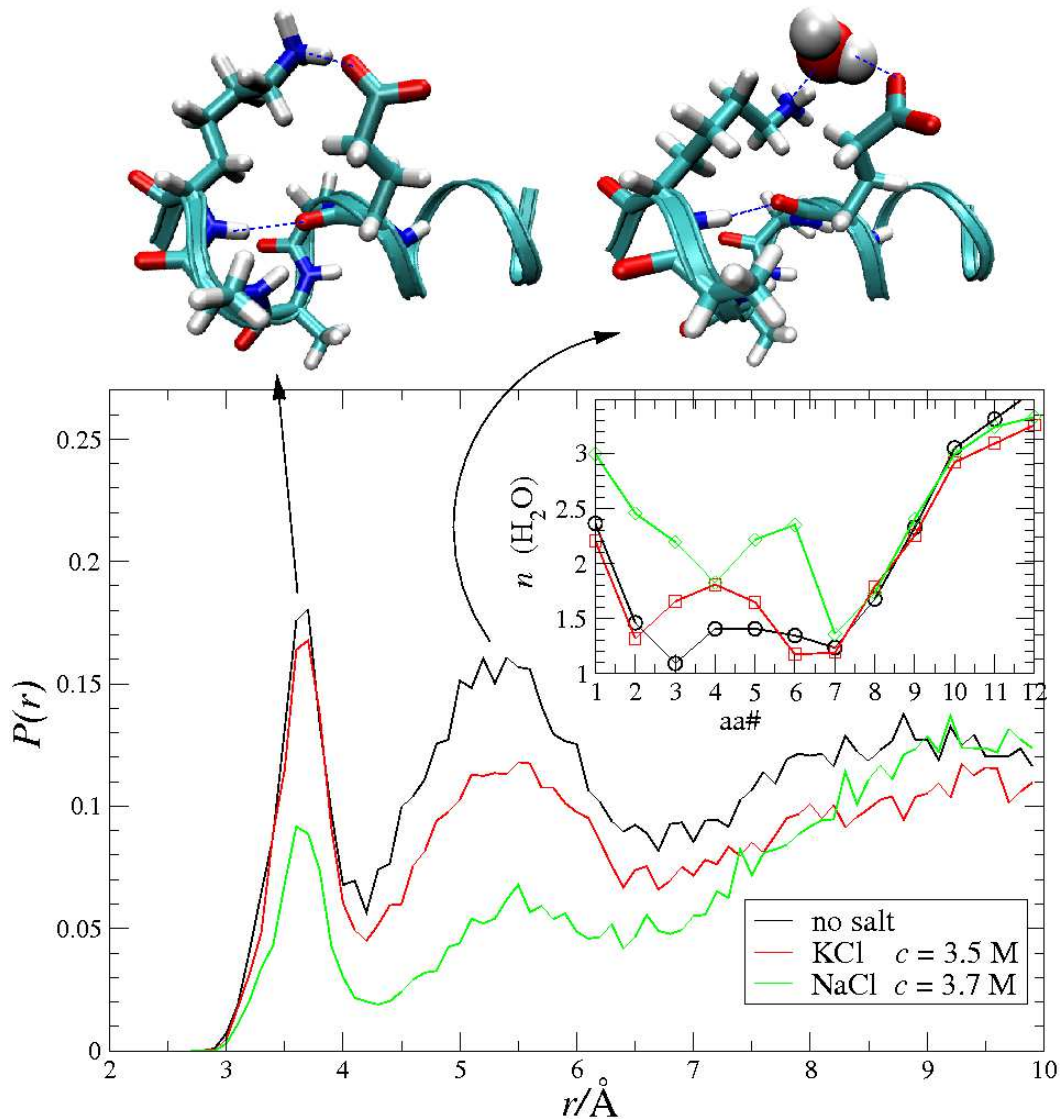}
\caption{Probability distribution of distances between the carboxylate carbon atom on the Glu7 side chain and the amine nitrogen on the Lys11 side chain. The MD snapshots on the top exemplify direct and indirect (separated by one water molecule) salt-bridge configurations corresponding to the first and second peak of $P(r)$, respectively. Just side chains of amino acids 7-11 are shown. Note the shielding of the Glu7 backbone carbonyl hydrogen bonding to the Lys11 nitrogen (blue dashed line). Inset: Average number of water oxygens in the first solvation shell of the backbone carbonyl atoms vs. amino acid number (aa \#). Lines are guide to the eye.}
\label{fig:1}
\end{figure}

\begin{figure}[htp]
\includegraphics[width=14cm,angle=0]{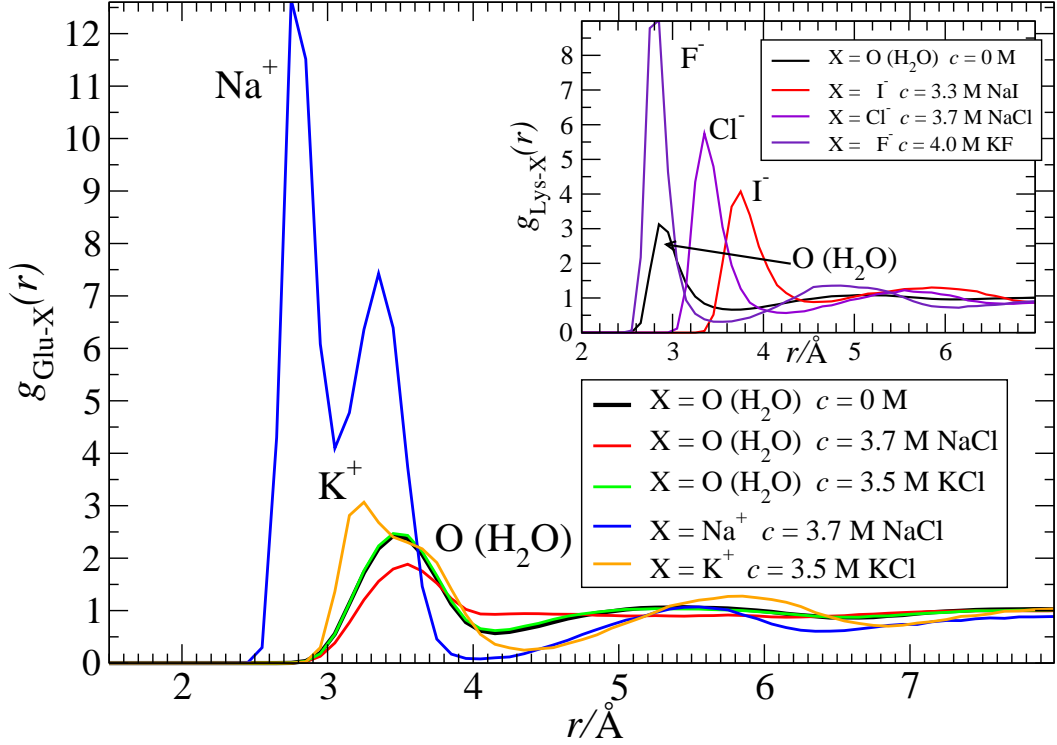}
\caption{Radial distribution function (rdf) $g_{\rm Glu-X}(r)$ between the carboxylate carbon of the Glu side chains and X, where X = O (water oxygen), Na$^{+}$, or K$^{+}$, plotted for no salt ($c=0$) and NaCl and KCl at concentrations $c\simeq 3-4$~M. Note that $g_{\rm Glu-O}(r)$ at $c=0$ (black line) and $g_{\rm Glu-K^{+}}(r)$ (green line) are nearly indistinguishable. Inset: rdf $g_{\rm Lys-X}(r)$ between the nitrogen of the Lys side chain and X, where X = O (water oxygen), F$^{-}$,Cl$^{-}$, or I$^{-}$. } 
\label{fig:1}
\end{figure}

\begin{figure}[htp]
\includegraphics[width=14cm,angle=0]{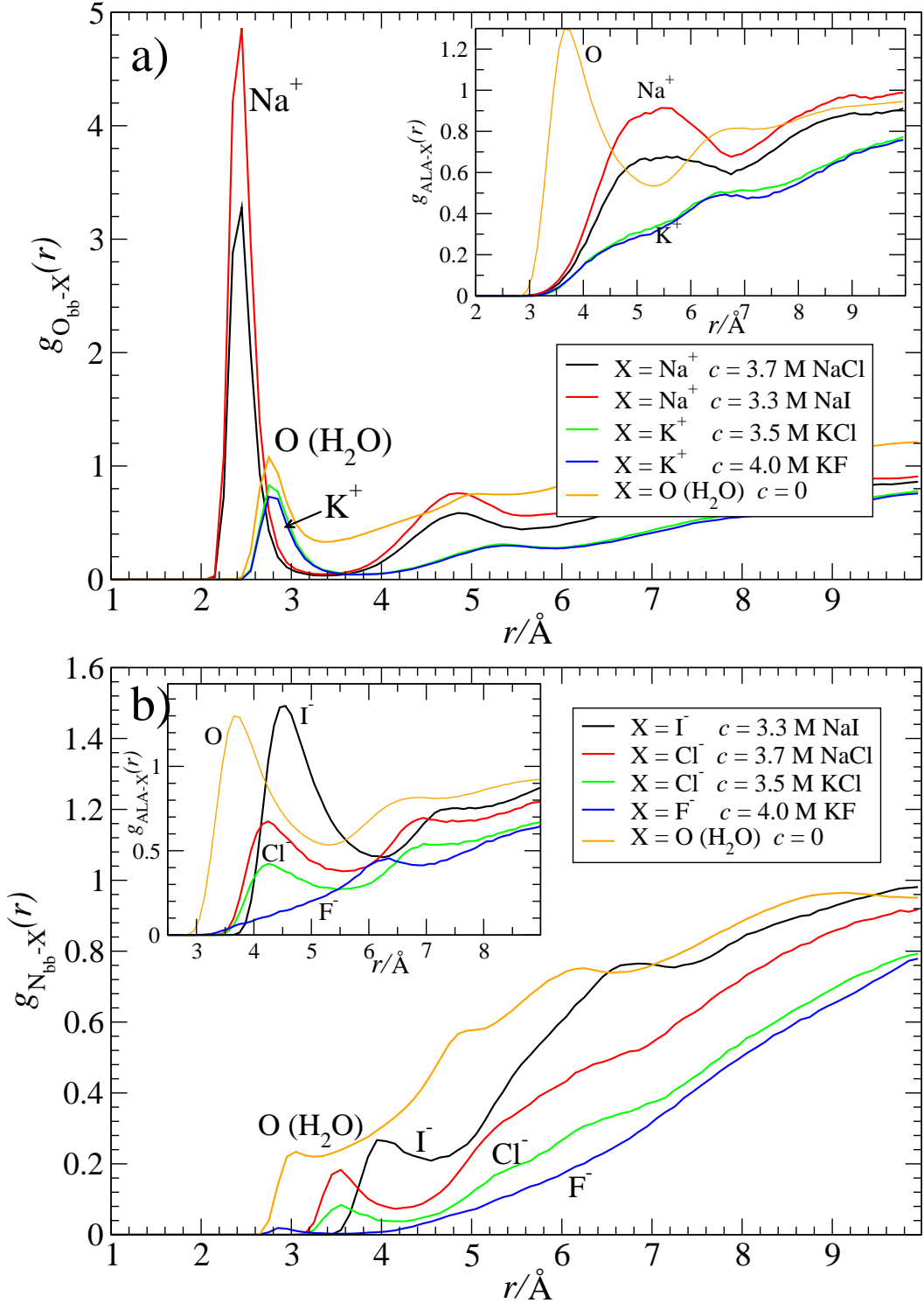}
\caption{a) Radial distribution function (rdf) $g_{\rm O_{bb}-X}(r)$ between the backbone  oxygen and X, where X = Na$^{+}$, K$^{+}$, or water oxygen (O) plotted for different salts, see legend. Inset: rdf $g_{\rm ALA-X}(r)$ between alanine side chains  and X. b) Rdf $g_{\rm N_{\rm bb}-X}(r)$ between the backbone nitrogen and X, where X = F$^{-}$,Cl$^{-}$,  I$^{-}$, or water oxygen (O) for different salts, see legend.  Inset: rdf $g_{\rm ALA-X}(r)$ between alanine side chains and X.} 
\label{fig:1}
\end{figure}

\begin{figure}[htp]
\includegraphics[width=14cm,angle=0]{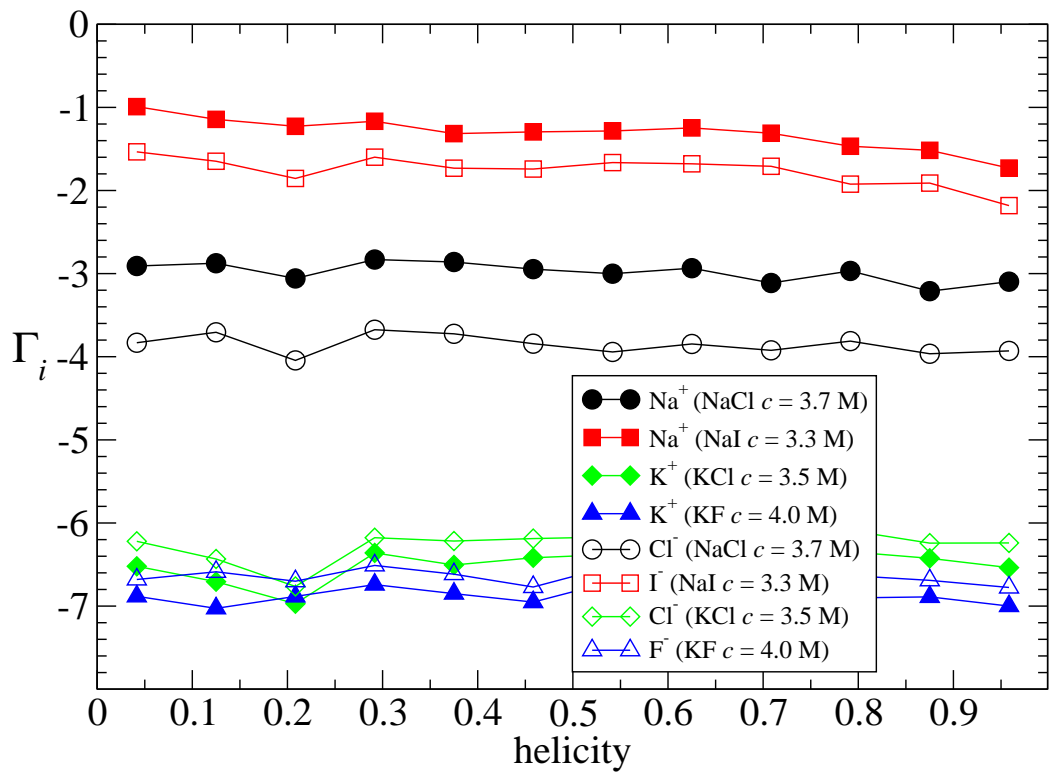}
\caption{Preferential interaction parameter $\Gamma_{i}$ for every ionic species $i$ in the first peptide solvation shell (averaged over every peptide atom) vs. peptide helicity.}
\label{fig:1}
\end{figure}

\begin{figure}[htp]
\includegraphics[width=14cm,angle=0]{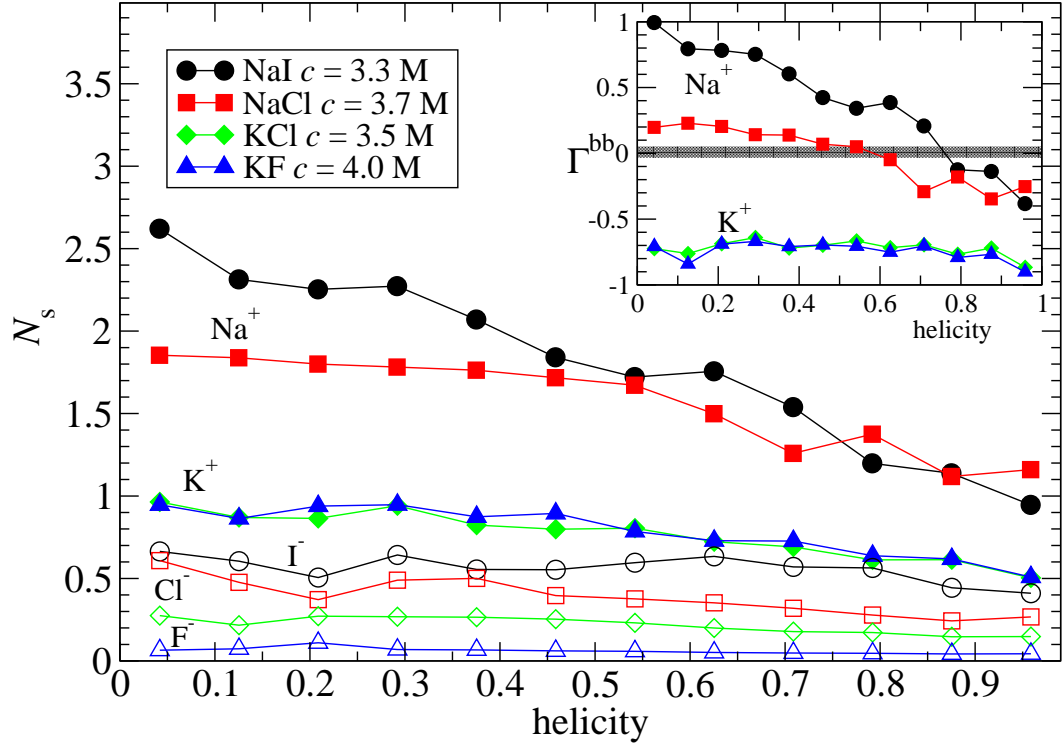}
\caption{Average number of cations (Na$^{+}$ and K$^{+}$, filled symbols) and anions (I$^{-}$, Cl$^{-}$, and F$^{-}$, open symbols) 
in the first solvation shell of the backbone amides vs. helicity for different salts at concentrations $c \simeq$ 3-4M. Inset: preferential interaction parameter $\Gamma_{bb}$ for Na$^{+}$ and K$^{+}$ only for the first backbone 
solvation shell vs. helicity.}
\label{fig:1}
\end{figure}
 
\begin{figure}[htp]
\includegraphics[width=14cm,angle=0]{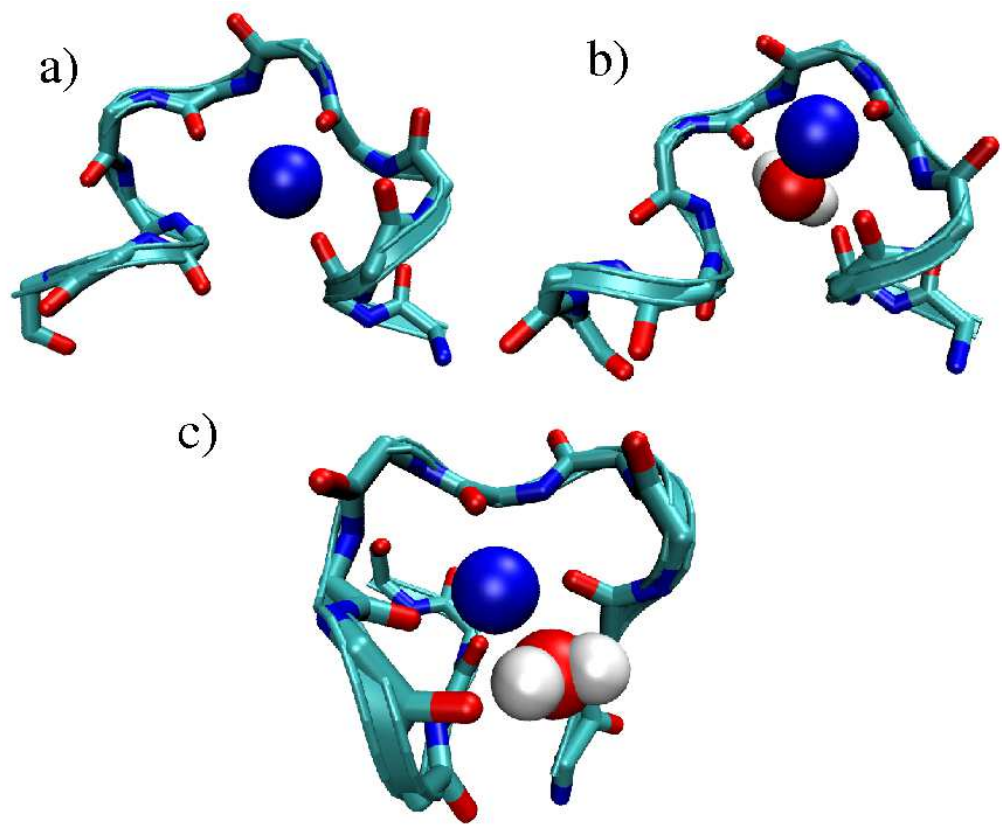}
\caption{MD simulations snapshots of long-lived ($\sim$10-20 ns) peptide configurations observed 
in NaCl or NaI  solutions at  large concentrations, where a), one Na$^{+}$-ion 
(blue sphere) or, b), c), Na$^{+}$ {\it and} one water molecule (red and white spheres) are permanently hydrogen bonded and immobilized by  the peptide backbone. Only backbone atoms of the peptide are shown.}
\label{fig:1}
\end{figure}

\clearpage
\newpage

\begin{figure}[htp]
\includegraphics[width=14cm,angle=0]{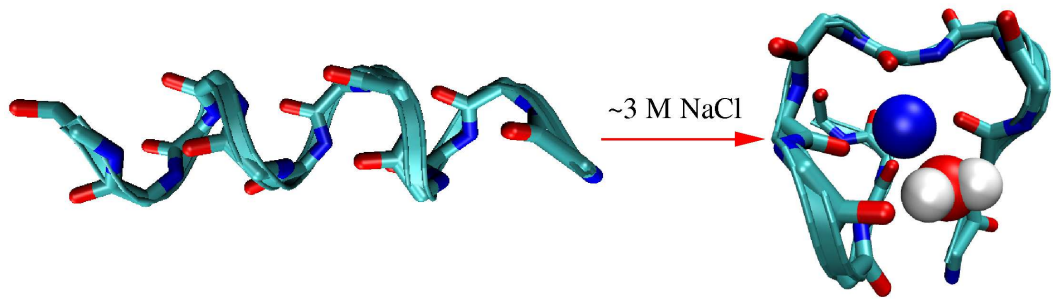}
\caption{TOC figure.}
\label{fig:1}
\end{figure}

\end{document}